\documentclass[manuscript,screen,acmsmall]{acmart}

\usepackage{algorithm}
\usepackage[noend]{algpseudocode}
\usepackage{xspace}
\usepackage{enumitem}
\usepackage{lineno}
\usepackage{array}
\usepackage{cleveref}

\usepackage{xcolor}

\usepackage{framed}
\definecolor{shadecolor}{gray}{0.91}

\usepackage{algpseudocode}
\usepackage{scalefnt}

\usepackage{wrapfig}

\usepackage{datetime2}

\setlist[description]{leftmargin=*,labelindent=*}

\newcommand{\alglinenoNew}[1]{\newcounter{ALG@line@#1}}
\newcommand{\alglinenoPop}[1]{\setcounter{ALG@line}{\value{ALG@line@#1}}}
\newcommand{\alglinenoPush}[1]{\setcounter{ALG@line@#1}{\value{ALG@line}}}

\newcommand{\mypara}[1]{\smallskip\noindent\textbf{#1.}}

\newcommand{\ia}{\textit{i}}
\newcommand{\ib}{\textit{ii}}
\newcommand{\ic}{\textit{iii}}

\newcommand{\com}[1]{}

\algdef{SE}[Receiving]{Receiving}{EndReceiving}[1]{\textbf{upon
		receiving}\ #1\ \algorithmicdo}{\algorithmicend\ \textbf{}}%
\algtext*{EndReceiving}

\algdef{SE}[Upon]{Upon}{EndUpon}[1]{\textbf{upon}\ #1\ \algorithmicdo}{\algorithmicend\ \textbf{}}%
\algtext*{EndUpon}

\newcommand\StateX{\Statex\hspace{\algorithmicindent}}
\newcommand\StateXX{\StateX\hspace{\algorithmicindent}}
\algrenewcommand\textproc{}

\makeatletter \let\sv@thm\@thm \def\@thm{\let\indent\relax\sv@thm} \makeatother

\newcommand{\calX}{\mathcal{X}}

\crefname{table}{table}{tables}
\crefname{table}{Table}{Tables}
\crefname{algocf}{alg.}{algs.}
\crefname{algocf}{Alg.}{Algs.}
\crefname{figure}{Fig.}{Figs.}
\crefname{figure}{fig.}{figs.}
\crefname{claim}{claim}{claims}
\crefname{claim}{Claim}{Claims}
\crefformat{chapter}{\S#2#1#3}
\crefmultiformat{chapter}{\S\S#2#1#3}{and~#2#1#3}{, #2#1#3}{, and~#2#1#3}

\crefformat{section}{\S#2#1#3}
\crefmultiformat{section}{\S\S#2#1#3}{and~#2#1#3}{, #2#1#3}{, and~#2#1#3}

\usepackage{enumitem}
\setlist{nosep} 
\setlist{itemsep=1pt, topsep=3pt}

\newcommand{\TL}{TL\xspace}

\newcommand{\WL}{WL\xspace}

\newcommand{\GSN}{Grassroots Social Networking\xspace}

\AtBeginDocument{%
  \providecommand\BibTeX{{%
    \normalfont B\kern-0.5em{\scshape i\kern-0.25em b}\kern-0.8em\TeX}}}

\setcopyright{acmlicensed}
\copyrightyear{2018}
\acmYear{2018}
\acmDOI{XXXXXXX.XXXXXXX}

\acmISBN{978-1-4503-XXXX-X/18/06}

\begin{document}

\title[Grassroots Social Networking]{Grassroots Social Networking:  Where People have Agency over their Personal Information and Social Graph}

\author{Ehud Shapiro}
\affiliation{%
 \institution{Weizmann Institute of Science}
   \country{Israel}}
   \affiliation{%
 \institution{London School of Economics}
   \country{United Kingdom}\\
   \\
}

\renewcommand{\shortauthors}{Shapiro}

\begin{abstract}
Offering an  architecture for social networking in which people have agency over  their personal information and social graph is an open challenge. Here\footnote{This paper is a revision and  extension of~\cite{shapiro2023gsn}} we present a grassroots architecture for serverless, permissionless, peer-to-peer social networks termed \textbf{Grassroots Social Networking} that aims to address this challenge. The architecture is geared for people with networked smartphones---roaming (address-changing) computing devices communicating over an unreliable network (e.g., using UDP).  The architecture incorporates (\ia) 
a decentralized social graph, where each person controls, maintains and stores only their local neighborhood in the graph; (\ib) personal feeds, with authors and followers who create and store the feeds; and (\ic) a  grassroots dissemination protocol, in which communication among people occurs only along the edges of their social graph. The architecture realizes these components using the \emph{blocklace} data structure -- a partially-ordered conflict-free counterpart of the  totally-ordered conflict-based blockchain.
We provide two example Grassroots Social Networking protocols---Twitter-like and WhatsApp-like---and address their security (safety, liveness and privacy),  spam/bot/deep-fake resistance, and implementation, demonstrating how server-based social networks could be supplanted by a grassroots architecture. 
\end{abstract}



\received{20 February 2007}
\received[revised]{12 March 2009}
\received[accepted]{5 June 2009}

\maketitle

\section{Introduction}

\subsection{Background} 
The architecture of server-based social networks---whether centralized (Facebook, X), federated~\cite{raman2019challenges}, or blockchain-based~\cite{DSNP}---results in the network operators obtaining and storing a broad range of personal information of their members, including their online behavior, original content, and social graph.
This ability is at the heart of surveillance capitalism~\cite{zuboff2023age}---the business model of all leading social networks---which perfected the commercial exploitation of personal information and original content with little or no remuneration, the provision of biased information in the service of commercial interests,  and the opaque governance of membership and content, also in the service of commercial interests.  In order to mitigate the damage wrecked by surveillance capitalism, governments attempted to legislate the storage and exploitation of personal information, notably the GDPR~\cite{regulation2016GDPR} and DSA~\cite{regulation2022DSA} regulations by the EU.
While this and other regulations have increased the care with which personal information is handled, it also resulted in people consenting a dozen times a day to surrender ownership and control of their personal information to web sites they visit and applications they use. More importantly,
it has not changed the fundamentals of the surveillance-capitalism business model of global social networking platforms.

Alternatives to centrally-controlled social networks have been explored for the last two decades.  These include federated architectures such as Mastodon~\cite{raman2019challenges}, peer-to-peer architectures such as distributed pub/sub systems~\cite{chockler2007constructing,chockler2007spidercast}, Gryphone~\cite{strom1998gryphon}, PeerSON~\cite{buchegger2009peerson}, and more~\cite{guidi2018managing,jiang2019bcosn}, and blockchain-based protocols such as the Decentralized Social Networking Protocol (DSNP)~\cite{DSNP} and others\cite{guidi2018managing, rahman2020blockchain}.  While all these proposals aim to subvert the centralization of power of global platforms, they still require members to surrender their personal information to third parties, whether an operator of a federated server, agents maintaining a distributed hash table or a distributed file system, or miners/validators of a blockchain protocol.
Secure Scuttlebutt (SSB)~\cite{tarr2019secure} is a peer-to-peer protocol, mesh network, and self-hosted social media platform in which each participating computer acts as a server. As Secure Scuttlebutt is the closest to our work in spirit and goal, we relate to it in Section \ref{section:scuttlebutt}.

But what if the social network could operate directly and solely on people's smartphones?  This may alleviate the need to surrender personal information to third parties altogether.  If people could participate in a social network while retaining ownership and control of their personal information, this would hollow out surveillance capitalism and open the door to new business models, e.g., data cooperatives~\cite{buhler2023unlocking},  in which members have full control of the use of their personal information and original digital content and can benefit directly and fairly from its commercial exploitation.

One may conceive of several challenges to serverless, smartphone-based social networks:
\begin{enumerate}
    \item \textbf{Performance:} Today's smartphones have hundreds and thousands of times the computing, memory and networking capabilities of the Unix workstations that were the workhorses  of the Internet at its inception.  Yet, they function today mostly as adorned terminals to services offered by cloud/server-based platforms  (with peer-to-peer audio and video communication as the exception).  So why aren't smartphones being used to their full capacity?  Why aren't there yet smartphone-based social networks? The answer should be sought for not in their performance limitations, but in other challenges such as those listed below.
    
    \item \textbf{Business Model:} Surveillance-capitalism is the cash-cow of global digital platforms. To make a dent in this business model, smartphone-based social networks have to offer alternatives that provide for a viable market without compromising people's ownership and control of their personal information the products of their digital labor.  For example allowing the fair trade in personal information and revenue sharing through data cooperatives and ad purchasing groups.  

    \item \textbf{Internet:} Today, the Internet is a hostile environment for smartphone peer-to-peer applications:  Smartphones have no easy way to find each other, and once they do they may have difficulties communicating with each other.   Internet servers are easily discovered via the Domain Name System (DNS) protocol.  Smartphones change their Internet Protocol (IP) address frequently, when switching the Internet connection between a WiFi and an Internet Service Provider (ISP) and when traveling and roaming.
    An extension of DNS  that supports dynamically-allocated Internet addresses, Dynamic DNS (DDNS), was defined early on~\cite{DDNS} but was never extended to provide the full functionality needed for smartphones to find each other, let alone deploy such an extension as part of the global Internet infrastructure. Hence, even though smartphones are uniquely identified by their MAC address and possibly also by their phone number, today the Internet does not include a globally-available service similar to DNS that allows smartphones to find each.   
    
    Furthermore, even if two smartphones discover each other's current IP address, they may not be able to communicate freely.  Normally, smartphones are hidden behind  Network Address Translation (NAT) servers and firewalls, that may or may not allow certain types of phone-to-phone communication to go through. 
    
    A rather elaborate set of protocols and infrastructure (named STUN, TURN, TURNS, ICE) is needed to overcome these Internet limitations.  Such an infrastructure is deployed, for example, by Meta/WhatsApp, which in return knows and owns the social graph and personal information regarding the online behavior of around 2Bn people worldwide.

    \item \textbf{System Architecture:} Due to all the challenges presented above, progress on smartphone-based peer-to-peer architectures and protocols for Internet applications in general and social networking in particular has been lagging.  Addressing this lack---in particular for social networking---is the subject of this paper.
\end{enumerate}

\subsection{\GSN Concepts and Architecture}

\GSN offers a grassroots architecture and protocol~\cite{shapiro2023grassroots,shapiro2023grassrootsBA} for serverless, permissionless, peer-to-peer social networks in which people retain ownership and control over their personal information, original digital content, and social graph. \GSN is geared for roaming personal computing devices, namely smartphones, and is guided by three fundamental concepts: 
\begin{enumerate}[leftmargin=*, partopsep=0pt,topsep=0pt,parsep=0pt]
    \item \textbf{Freedom of Digital Assembly}~\cite{mill1859liberty}, the ability of people to assemble freely digitally, not under the control of, and without surrendering personal information to, any third party.
    \item  \textbf{Digital Sovereignty}~\cite{pohle2021digital}, the ability of people to conduct their social, economic, civic, and political lives in the digital realm solely using the networked computing devices they own and operate (e.g., smartphones), free of third-party control, surveillance, manipulation, coercion, or rent seeking.  
    \item \textbf{Grassroots Systems}~\cite{shapiro2023grassroots,shapiro2023grassrootsBA}, which are permissionless distributed system that can have multiple oblivious instances (independent of each other and of any global resources)  that can interoperate once interconnected.  \GSN are an example of grassroots systems: Serverless peer-to-peer smartphone-based social networks supporting multiple independently-budding communities that may merge when a member of one community becomes also a member of another. We note that most systems in the digital realm are not grassroots, including centrally-controlled digital platforms (Facebook) and distributed systems designed to have a single global instance (Bitcoin~\cite{bitcoin}, Ethereum~\cite{buterin2014next}) or to employ a single replicated (Blockchain~\cite{bitcoin}), or distributed (IPFS~\cite{benet2014ipfs}, DHT~\cite{rhea2005opendht}) shared global data structure, and distributed pub/sub systems with a global directory~\cite{chockler2007constructing,chockler2007spidercast,buchegger2009peerson}).    
\end{enumerate}

With these concepts we present the architecture of Grassroots Social Networking (GSN) that supports digital freedom of assembly and digital sovereignty
for people with smartphones. We provide two example Grassroots Social Networking protocols: Twitter-like (which is also LinkedIn- and Instagram-like), and WhatsApp-like, demonstrating how centrally-controlled social networks could be supplanted by a serverless, permissionless, peer-to-peer grassroots architecture. 

The \GSN architcture has three components:
\begin{enumerate}[leftmargin=*]
    \item \textbf{Decentralized\footnote{By \emph{decentralized} we mean decentralized control,  so that each agent controls only part of the data structure. Contrast this with standard blockchain applications, where all miners vie for control of the blockchain's single tip.} Social Graph}, with each member controlling, maintaining and storing its local neighbourhood in the graph.  
    \item \textbf{Member-Created Feeds}, each with authors and followers.
    \item \textbf{Grassroots Dissemination}~\cite{shapiro2023grassroots}, a peer-to-peer protocol carried over the edges of the social graph, where friends update each other with information they know and believe their friends need.
\end{enumerate} 

The GSN architecture realizes these components using the \textbf{blocklace} data structure~\cite{shapiro2023grassroots,keidar2022cordial,lewispye2023flash} -- the distributed\footnote{A data structure is \emph{distributed} if different agents access, maintain and store different parts of the data structure.  It is \emph{replicated} if all agents must obtain and maintain a complete copy of the entire data structure in order to operate, as is the norm in blockchain applications.  Note that the term  `distributed ledger' is often used to refer to a replicated ledger.}  partially-ordered counterpart of the replicated totally-ordered blockchain. The blocklace uses blocks to record the evolving social graph; to immutably record acts and their causal relations, supporting members in creating and following feeds; and to function as multichannel ack/nacks, supporting dissemination over an unreliable network.

The Twitter-like network has \emph{public feeds}, each with a sole author and permissionless followers. The WhatsApp-like network has \emph{private groups}, members of each group being both the authors and followers of its feed. Naturally, an actual Grassroots Social Networking application may integrate the two protocols to provide both public feeds and private groups.

\subsection{Security of \GSN} 

Security concerns can be separated into  integrity/safety, availability/liveness, and confidentiality/privacy.
We address each of them in turn.

\mypara{Safety} In existing social networks, utterances by members do not carry intrinsic attribution or provenance.  Hence,
 Twitter members must trust the social network operator to correctly identify the authorship of a tweet. As a screenshot with a tweet could easily be a fake, an author of a tweet can later repudiate it simply by deleting it, and no proof that the tweet ever existed can be provided, except perhaps by the operator itself.
In WhatsApp, members of a group must trust the service to correctly identify the member who authored an utterance, and utterances forwarded from one group to another have no credible attribution or context. 

Thus, the \emph{safety requirement of \GSN} is that any utterance of a member can be attributed to that member, and in the correct context in which it was uttered.  This safety requirement is realized by each utterance of a member $p$ being embedded as the payload of a block $b$ created and signed by $p$, which also includes hash pointers to the blocks that are causally-precedent to $b$.   Thus, an utterance in a payload of a block can be attributed to the signatory of the block, and can be placed in the context of the utterances in the blocks that causally precede it.  An utterance can be forwarded with correct attribution and context either by forwarding a pointer to the block or by including the entire block as a payload of a new block.  In either case, the author of the original utterance as well as its context can be correctly attributed.  A consequence of doing so is that  utterances that have been forwarded multiple times are provided with their entire provenance.

\GSN safety implies that each member can be held accountable to their own utterances, as well as to utterances that they forward. As \GSN has no controlling entity that can be held accountable in case members of the social network break the law,  accountability and law enforcement in \GSN are more similar to real life then to the social networks we are familiar with.  In particular, the  means needed to break into  an illegal digital organization that uses \GSN would be similar to those needed to break into any illegal organization:  Snitches, undercover agents, wiretapping, etc., as there is no single third party that can be coerced into uncovering or blocking the digital organization.

An equivocation consists of two  blocks authored by the same person, intentionally unrelated in the blockchain/blocklace and intentionally distributed to different people.  Equivocations are excluded by payment systems~\cite{bitcoin,buterin2014next,guerraoui2019consensus,lewispye2023flash, lewis2023grassroots}, as they may result in so-called `double-spending'.  However, their exclusion is not required in social networking, and hence social networking protocols can be simpler than payment systems protocols.  In particular, using a blockchain protocol to realize social networking~\cite{DSNP} may be an overkill with unwarranted costs.  

While the \GSN protocol does not exclude equivocations, thanks to its liveness---discussed next---both equivocating utterances to a group are eventually observed by all correct members of the group, who would then identify the culprit and can respond with social means.  This property, while not sufficient for the safety of payment systems,  seems natural and sufficient for social networking.

\mypara{Liveness}
\GSN liveness theorems state conditions under which a block issued by an author to a feed eventually reaches a follower of this feed.  For  Twitter-like feeds, the liveness condition is that  the follower is connected to the author via a path in the social graph that consists of correct members, all of whom follow the author.  For the WhatsApp-like groups, the liveness condition is that the follower and an agent that knows the block are correct members of the group.

In describing the protocols, we opted for the simplest and minimal protocols that satisfy safety and liveness, forgoing possible optimizations of latency or complexity. 

\mypara{Privacy} As each member is equipped with a keypair of its choosing and is identified by their public key, privacy can be ensured by encryption. Privacy of direct messaging is realized by encrypting messages using the public key of the intended recipient.  Privacy of a group is realized by its founder creating a unique keypair to encrypt communication within the group, and secretly sharing it with group members using their public keys.  

Clearly, members can copy any text they wish and paste it anywhere they wish.  In addition,  an agent  may copy an entire block and place it as a payload of another block and send it to others.  This will not only include the text, but also the signature of its author, as well as pointers to other blocks of the blocklace encoding the conversation of that group.  Even if the text is encrypted (e.g. with the group's key as in the WhatsApp-like protocol),  nothing prevents the sharing member from also sharing the decryption key.  As in any other setting, anyone who knows a key might share it with anyone they please, and anyone might breach confidence and privacy.   However, since every block in GSN is signed,  when one breaches privacy within the protocol the breach carries their signature so the culprit can be identified.  

\mypara{Spam, bots and Deep Fake}
The GSN architecture is particularly resistant to spam, bots, and deep fake.  Since each block is signed and every forwarded block has provenance, anything without proper provenance can be filtered as spam (or implicate its creator/forwarder).  In particular, a deep-fake payload that is not attributed to its source can be promptly filtered as spam, so that members are expected to never be exposed to bots or deep fake unless they choose to.  

We note that in a centrally-controlled social network, the operator may know every utterance by every member, and
can use an algorithm that best serves its own commercial or other interests in order to determine the information a member sees, and in what order. 
Grassroots social networks let each member know everything they need based on their social graph.  Furthermore, each member can filter and display the information received in away that best serves their need using any client app that follows the protocol,  including an app that lets the member customize the rules that prioritize and filter the displayed information, manually or using computational agents of its choosing, including AI.

\mypara{Byzantines and Sybils} A Byzantine agent cannot fake a block by another agent as long as the latter's private key has not been compromised.  If a key is compromised, their holder can notify all their friends of this using any available means, including through the protocol itself if they have not been disconnected.  They can then provide their friends with their newly-created public key. 
Byzantine agents can also equivocate but, as discussed, the damage this creates in social networks is limited, and eventually they will be held accountable for such a fault.

Multiple identities of a person could be useful in the context of social networking, as they may serve the social needs of the person holding them without harming others, when used in separate social contexts.  As people choose their friends and whom to follow and obtain information only from them, Sybils may have limited success in this context, and if they cause any harm it would be primarily to the reputation of the person operating them.

\mypara{Implementation} The pseudocode provided for the protocols (Algs. \ref{alg:blocklace}, \ref{alg:tl}, \ref{alg:wl}) is readily implementable.  The key technological challenge is how to let smartphones find each other and then communicate.  If a person's smartphone knows its own IP address, it can communicate it to their friends' smartphones using any available means (e.g. SMS), to bootstrap an initial connection or to reestablish it in case it was completely lost.  Once phones are connected, the protocols aim to keep the connection even when smartphones change their IP address or are temporarily unavailable.  This functionality can be provided by  the WebRTC/ICE framework.  However, at present, avoiding the use of globally-addressable servers may not always be possible due to the vagaries of firewalls and network address translation (NAT): If a smartphone is located behind a NAT, it might need to use a publicly-available STUN server to identify its global IP address;  and if located behind a firewall that prevents smartphones from communicating directly, it would need a publicly-available TURN/TURNS server to relay the communication.

\subsection{A Grassroots Architecture for the Digital Realm}

\GSN was developed as part of a broader vision of a grassroots architecture for the digital realm~\cite{shapiro2022foundations},  outlined here. The digital realm today is dominated by autocratic  (1 person -- all votes) global digital platforms such as Facebook, Twitter and the like.  Federated systems aim to distribute control, but federated servers are still controlled autocratically.
Blockchain-based cryptocurrencies such as Bitcoin and Ethereum and their DeFi applications are global platforms with decentralized control.  However, their control is intrinsically plutocratic (1 coin -- 1 vote).  Looking for systems that offer egalitarian control and democratic governance (1 person -- 1 vote) in an intrinsic way, namely not at the courtesy of an underling autocratic or plutocratic platform, we find none (but see the discussion of Secure Scuttlebutt in Section \ref{section:scuttlebutt}).

We contend that egalitarian and democratic systems cannot emerge without a suitable architecture, and
our long-term goal is to develop such.  Our current proposal and vision of a grassroots architecture is presented in reference~\cite{shapiro2024grassroots}. 

\GSN, the subject of this paper is a key component in the grassroots systems architecture and an enabler of higher-level applications.  A key among them is grassroots cryptocurrencies~\cite{shapiro2022gc}, which are a means for turning mutual trust into liquidity,  and thus offer a novel route to ``banking the unbanked''\cite{agarwal2017banking,dupas2018banking,bruhn2009economic}.
The goal of grassroots cryptocurrencies is to provide a foundation with which local digital economies can emerge independently of global digital platforms and of global cryptocurrencies; can form and grow without initial capital or external credit; can trade with each other; and can gradually merge into a global digital economy.  

Note that \GSN requires dissemination, and Grassroots Cryptocurrencies require also equivocation exclusion, but neither require consensus.  Consensus is needed only  by higher-level applications that employ ``on-chain'' governance such as grassroots social contracts~\cite{cardelli2020digital}, the grassroots counterpart of miner-operated smart contracts, and grassroots representative assemblies (in preparation).

Formally proving that a system is grassroots is an endeavor that requires significant mathematical machinery~\cite{shapiro2023grassroots,shapiro2023grassrootsBA,lewis2023grassroots}.  Rather than doing so here, we only make an informal claim that \GSN is indeed grassroots.  We rely on the fact that Grassroots Dissemination has been proven to be grassroots previously~\cite{shapiro2023grassroots}, and that this is the only communication protocol employed by \GSN.  In particular, the Twitter-like protocol presented here has the same notion of social graph, follow and friends as the formal Grassroots Dissemination protocol of~\cite{shapiro2023grassroots}, and the WhatsApp-like protocol can be reduced to the formal Grassroots Dissemination protocol by mapping every WhatsApp group (represented as a hyperedge, see below) into a directed clique of its members.

\mypara{Roadmap} The rest of the paper is organized as follows:  Section \ref{section:GSN} present the \GSN architecture; Section  \ref{section:TL} presents the Twitter-like protocol \TL; Section \ref{section:WL} presents the WhatsApp-like protocol \WL; Section \ref{section:scuttlebutt} compares \GSN with Secure Scuttlebutt; and Section \ref{section:conclusions} concludes.  

\section{\GSN Architecture}\label{section:GSN}

\GSN is based on two principles:

\vspace{0.5em}
\fcolorbox{gray!100}{gray!5}{%
    \parbox{0.96\textwidth}{%
\textbf{Principles of \GSN:}
\begin{enumerate}[leftmargin=*]
    \item \textbf{Friendship}: Make friends by mutual consent. 
    \item \textbf{Grassroots Dissemination}: Tell to your friends things you know and believe they need.
\end{enumerate}
        }
    }
\vspace{0.5em}
\newline
\GSN realizes these principles using three components:

\vspace{0.5em}
\fcolorbox{gray!100}{gray!5}{%
    \parbox{0.96\textwidth}{%
    \textbf{Components of \GSN:}
\begin{enumerate}[leftmargin=*]
    \item \textbf{Decentralized Social Graph}, with members controlling, maintaining, and storing their social neighbourhoods in the graph,  e.g., own vertex and adjacent edges.
    \item \textbf{Member-Created Feeds},  with authors and followers.
    \item \textbf{Grassroots Dissemination}, a protocol for communication along the edges of the social graph, through which friends update each other with information they know and believe their friends need.
\end{enumerate}
     }
    }
    \vspace{0.5em}
\newline
All these components are realized via the blocklace, as follows:

\vspace{0.5em}
\fcolorbox{gray!100}{gray!5}{%
    \parbox{0.96\textwidth}{%
\textbf{In \GSN, the blocklace:}
\begin{enumerate}[leftmargin=*]
    \item \textbf{Social Graph}: Encodes the evolving decentralized social graph, allowing members to create, maintain and store their neighbourhood in the graph.
    \item \textbf{Feeds}:  Immutably records utterances and their causal relations for members to create and follow feeds.
    \item \textbf{Dissemination}: Provides multichannel ack/nacks for grassroots dissemination over an unreliable network.
\end{enumerate}
   }
    }

\vspace{0.5em}
We elaborate on each of these components in turn.

\begin{figure}
  \center{\includegraphics[width=\textwidth]{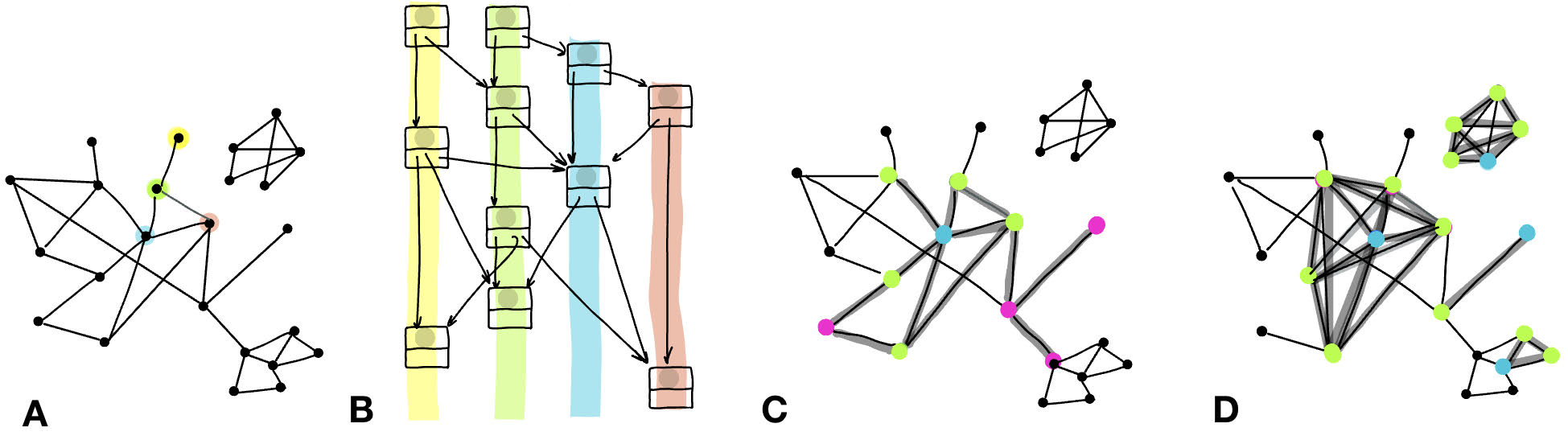}}
  \caption{\textbf{Grassroots Social Networks.} \textbf{A}. A Social graph, with some colored vertices. \textbf{B}. A blocklace with interlinked color-coded personal blockchains.  \textbf{C}. A Twitter-like social graph, with a feed (blue) friends (green) and followers (pink). \textbf{D}. A WhatsApp-like social graph, with groups of varying sizes, each with a founder (blue) and members (green).  A person can be a founder and a member of multiple groups (color-coding not shown).}
  \label{fig:teaser}
\end{figure}

The social networking protocols presented here engage in the grassroots formation of a social graph.  This implies that at any point in time the social graph created may consist of disconnected components (See Fig. \ref{fig:teaser}). Even if subsequently the graph becomes connected, a new disconnected component may emerge at any moment.  In principle, each connected component of the graph may be searchable by its members via a cooperative distributed search protocol.  However, a social networking protocol cannot both be grassroots and support a global directory of all its members~\cite{shapiro2023grassroots}.

\subsection{The Social Blocklace} \label{sec:transactionBlocklace}
The Grassroots Social Networking architecture employs the blocklace to create and maintain the social graph, to allow members to create and follow feeds, and to disseminate information~\cite{shapiro2023grassroots}.  Basic notions of the blocklace, adapted for Grassroots Social Networking, are specified by Alg. \ref{alg:blocklace} and presented concisely herein.  

\begin{figure}[t]
\begin{algorithm}[H]
    \caption{\textbf{Social Blocklace Utilities.} Code for agent $p $}
    \label{alg:blocklace}
    \small
    \begin{algorithmic}[1] \scalefont{0.9}
        \Statex \textbf{Local variables:}

        \StateX struct $\textit{block } b$: \Comment{The structure of a block $b$ in a blocklace}
        \StateXX $b.\textit{id}$ -- block identifier, a hash pointer to the rest of the block signed by its creator $p$.  
        \StateX \Comment{We use $b.\textit{creator}$ or $\textit{id}.\textit{creator}$ for the signatory of  $b.\textit{id}$}
        \StateXX $b.\textit{address}$ -- IP address of $p$ at the time of creating the block
        \StateXX $b.\textit{payload}$ -- the block's payload
        \StateXX $b.\textit{pointers}$ -- a possibly-empty set of identifiers (signed hash pointers) of preceding blocks

 		\vspace{0.3em}
		\Procedure{$\textit{new\_block}$}{$\textit{payload}, \textit{pointers}$}   
		\State \textbf{new} $b$  \Comment{Allocate a new block structure}
        \StateX $b.\textit{address} \gets$  current IP address of $p$ 
	    \State $b.\textit{payload} \gets \textit{payload}$ 
        \State $b.\textit{pointers} \gets \textit{pointers}$
        \State $b.\textit{id} \gets \textit{hash}((b.\textit{address}, b.\textit{pointers}, b.\textit{payload}))$ signed by $p$
        \State \Return $b$
        \EndProcedure

         \vspace{0.3em}
        \Procedure{$\textit{observes}(b, b'', B)$}{} 
        \State \Return $b=b''~ \vee~  \exists \textit{id} \in b.\textit{pointers}  \wedge \exists b'  \in B \colon b'=(\textit{id},x,H) \wedge \textit{observes}(b', b'',B) $
        \EndProcedure

         \vspace{0.3em}
        \Procedure{$\textit{tips}$}{$B$} 
        \State \Return $\{b \in B : \not\exists b'\ne b \in B \wedge \textit{observes}(b',b)\}$
        \EndProcedure

         \vspace{0.3em}
        \Procedure{$\textit{ip\_address}(q, B)$}{} 
        \State \Return $\textit{address}$ s.t. $b\in B 
        \wedge ((b.\textit{creator} = q  \wedge address = b.\textit{address}) \vee b.\textit{payload} = (\textsc{IP},q,\textit{address})) \wedge \newline
        (\forall b'\in B : b'.\textit{creator} = q \implies \textit{observes}(b,b'))$
        \EndProcedure

        \alglinenoNew{counter}
        \alglinenoPush{counter}
 
    \end{algorithmic}
\end{algorithm}
\vspace{-2.5em}
\end{figure}

We assume a potentially-infinite set of agents $\Pi$ (think of all the agents yet to be produced), but when referring to a subset of the agents $P \subseteq\Pi$ we assume $P$ to be finite.
Each agent is associated with a single and unique key-pair of its own choosing, and is identified by its public key $p \in \Pi$. 
In practice, an agent is an app running on a smartphone, and hence a person may operate several agents by owning several smartphones. We discussed above the use of multiple identities and the implications of Sybils.

We also assume a collision-free cryptographic hash function \textit{hash} and a protocol-specific set of \emph{payloads} $\calX$ that includes the empty payload $\bot$.

A \emph{block} $b=(\textit{id},\textit{address},x,H)$ created by an agent $p$, also referred to as a \emph{$p$-block}, has an IP \textit{address}, a payload $x\in \calX$, a finite set $H$ of hash pointers to other blocks, and a signed hash pointer $\textit{id} := \textit{hash}((\textit{address},x,H))$ signed by $p$ (the signature also incorporates the public key $p$).
If the payload $x=\bot$ is empty then $b$ is an \emph{ack block}.
A hash-pointer in $H$ to a $p$-block is referred to as a \emph{self-pointer}, and if $H=\emptyset$ is empty then $b$ is \emph{initial}.
A \emph{blocklace} is a set of blocks (See Fig. \ref{fig:teaser}.A).
The following notions relate to a given blocklace $B$.  
A block $b\in B$ is a \emph{tip} of $B$ if no other block in $B$ points to $b$. 
A \emph{path} in $B$ from $b$ to $b'$ is a sequence of blocks in $B$, starting with $b$ and ending with $b'$, each with a pointer to the next block in the path, if any.  The path is \emph{empty} if $b=b'$.
A block $b$  \emph{observes} a block $b'$ in $B$ if there is a (possibly empty) path from $b$ to $b'$ in $B$. 

The \emph{closure} of a block $b$, denoted $[b]$, is the set of blocks $b$ observes, and the closure of a blocklace $B$ is defined by $[B]=\bigcup_{b\in B}[b]$.
A blocklace $B$ is \emph{closed} if it includes the closure of its blocks, namely $B=[B]$. 
The \emph{self-closure} of a $p$-block $b$, denoted $[b]_p$, are  the blocks reachable from $b$ via self-pointers.
A blocklace $B$ is \emph{self-closed} if it includes the self-closure of its blocks, namely $[b]_p\subseteq B$ for every $p \in \Pi$ and every $p$-block $b\in B$.

As we shall see, members of \TL maintain a single self-closed blocklace with blocks of the agents they follow, and members of \WL maintain a closed blocklace for each group they are a member of.


\subsection{The Social Graph}

The social graph of the  \GSN architecture has agents as vertices, but the type and meaning of the edges connecting them, the conditions defining when agents are friends, and how agents may update the social graph are all application specific.  
\begin{table}[!ht]\label{table:social-graphs}
  \caption{Social Graphs for Grassroots Social Networking and Grassroots Cryptocurrencies}
  \label{tab:commands}
  \begin{center}
  \smaller 
 \begin{tabular}{ | m{7em} | m{9em}| m{9em} | m{10em} | } 
    \hline
    \textbf{Protocol/\newline Social Graph} & \textbf{Grassroots TL} \textbf{(Twitter-Like)}
 & \textbf{Grassroots WL (WhatsApp-Like)}  & \textbf{Grassroots} \textbf{Cryptocurrencies}~\cite{lewis2023grassroots,shapiro2022gc}
\\
     \hline
    \hline
    \textbf{Graph type} & Directed  & Hypergraph 
                                                    & Undirected  \\
     \hline
    \textbf{Edge meaning} & Source agent follows destination agent
     & A group of agents  & Mutual disclosure of personal blockchains\\
 \hline
    \textbf{Friendship \newline condition} & Mutual following & Membership in same group & Mutual friendship offers
\\
    \hline

    \textbf{Graph update}  & Agent follows/unfollows another agent & Agent founds group;  invites/removes agents & Agents make/break friendships
\\
 \hline 
  \textbf{Agent needs}  &  Feeds of agents it follows &  Feeds of groups it is a member of
   &  Payments to it/in its coin; approvals of payments by/to it.\\
    \hline 
  \textbf{Data Structure }  &  Blocklace &  Blocklace   &  Blocklace
\\
 \hline   
  \end{tabular}
  \end{center}
\end{table}

The grassroots Twitter-like protocol \TL employs a directed graph.  A directed edge $p \rightarrow q$ from agent $p$ to agent $q$ means that $p$ \emph{follows} $q$. Any agent can create and remove outgoing edges at will.  We say that $p$ and $q$ are \emph{friends} if the social graph has edges in both directions, $p \leftrightarrow q$.
The grassroots WhatsApp-like protocol \WL employs a hypergraph (a graph in which an edge may connect any number of vertices).  A hyperedge connecting agents $P\subset \Pi$ means that the agents in $P$ are members in a \emph{group} represented by the hyperedge.  
We say that $p$ and $q$ are \emph{friends} if they are members of the same hyperedge of the social graph.
Any agent $p$ can create a group with $p$ as its sole member. A group creator can invite other agents to become members and remove members at will.  An invited agent may join the group and leave it at will.

A social graph is encoded by the blocklace using \textsc{reserved words} (quoted strings) in the payload.

The \TL directed graph encodes the creation of an edge $p\rightarrow q$ by a $p$-block with the payload $(\textsc{follow},q)$, and its removal by a $p$-block with the payload $(\textsc{unfollow},q)$.

In \WL, each hyperedge/group is realized by an independent blocklace, with blocks encoding group operations.
A newly created $p$-block with a payload as specified encodes an operation as follows:
\begin{enumerate}
    \item $(\textsc{group},\textit{name})$ -- create a new group named \textit{name} that corresponds to a new hyperedge in the social graph, labeled with the block's identifier and with $p$ as its founder and sole member.
    \item $(\textsc{invite},q)$ -- invite agent $q$ to a group 
    \item $\textsc{accept}$  -- accept an invitation and join the group
    \item $(\textsc{remove},q)$ -- remove agent $q$ from a group 
    \item $\textsc{leave}$ -- leave the group
\end{enumerate}
We note that, as in WhatsApp, private direct messaging can be realized by a two-member group.
We also note that in WhatsApp the founder can upgrade a member to be a manager, who can then invite and remove members. This extension can be incorporated here just as easily.

Both protocols use $(\textsc{ip},q,\textit{address})$ to record the current global IP \textit{address} of $q$.
We show below how the protocol preserves group integrity, provided the group creator is correct.

\subsection{Member-Created Feeds}

Member-created feeds are also  encoded by the blocklace using reserved words in payloads.
We employ two types of \emph{utterance blocks}, where an utterance of agent $p$ is encoded by a $p$-block with the following payload:
\begin{enumerate}
    \item $(\textsc{say},x)$, where $x$ is a string
    \item $(\textsc{respond},x,\textit{id})$, where $x$ is a string and $\textit{id}$ is an identifier (signed hash pointer) of an utterance block.
\end{enumerate}

In \TL, the utterances of a correct agent are incorporated in their personal feed.  Agents may choose to echo in their personal feed responses to their utterances by agents they follow,  thus making these responses available to all their followers.  As responses are signed, their author can be identified even when echoed.
In \WL, any utterance to a group by a correct group member is incorporated in the group feed.

\subsection{Grassroots Dissemination}

Grassroots dissemination employs the following principle:

\vspace{0.5em}
\fcolorbox{gray!100}{gray!5}{%
    \parbox{0.96\textwidth}{%
\textbf{Principle of Grassroots Dissemination:}\newline
An agent that needs a block can obtain it from a friend that has it.
   }
    }
    \vspace{0.5em}
\newline

This principle  is realized by each agent $p$ incorporating in every new $p$-block hash pointers to the current tips of $p$'s local blocklace.  Thus the block serves as a multichannel ack/nack:  It ack's blocks known to $p$ so far, and by the same token nack's blocks not yet known to $p$.   

In all protocols, every agent sends to every friend `every' block it produces (`every' is context-specific).
For example, in \TL, every agent $p$ sends to every friend $q$ every $p$-block produced within this protocol. A $p$-block informs of the most recent block known to $p$ by every agent $p$ follows, for example $q'$.  In such a case, if  $q$ happens to also follow $q'$, it may disseminate to $p$ any $q'$-block known to $q$ but, according to the most recent $p$-block received by $q$, is not yet known to $p$.

Similarly, in \WL, every agent $p$ sends to every member of a group $q$ every $p$-block uttered to that group.
If $q$ knows a block $b$ in the group feed that is not yet known to $p$ according to its most-recently received $p$-block, then $q$ may disseminate $b$ to $p$.

\section{\TL: A Grassroots Twitter-Like Protocol} \label{section:TL}

Here we present a Twitter-like protocol termed \TL, within the Grassroots Social Networking architecture.  The protocol employs blocks with the following payload keywords:
\begin{enumerate}
    \item \textsc{follow}, to add a directed edge to the social graph,\footnote{The handling of \textsc{unfollow} is deferred; it is not conceptually difficult but will clutter the exposition.}
    \item \textsc{say}, to `tweet',  
    \item \textsc{respond}, to respond to a tweet or a response, and
    \item \textsc{ack}, to acknowledge receipt of a block and disclose the blocks one knows
\end{enumerate}  

The \TL protocol realizes the principles and components of Grassroots Social Networking thus:

\vspace{0.5em}
\fcolorbox{gray!100}{gray!5}{%
    \parbox{0.96\textwidth}{%
\textbf{\TL Realization of \GSN}\\ by each agent $p$ with local blocklace $B$
\begin{enumerate}[leftmargin=*]
    \item \textbf{Friendship}: $p$ follows $q$ and/or accepts a friendship offer from $q$ by creating a $(\textsc{follow}, q)$ block. $p$ makes a friendship offer to $q$ by sending this block to $q$.
    \item \textbf{Decentralized Social Graph}: An edge $p\rightarrow q$ is added to the social graph by $p$ creating a $(\textsc{follow}, q)$ block. $B$  stores the immediate neighbourhood of $p$ in the graph,  including any $(\textsc{follow}, p)$ block received and $(\textsc{follow}, q)$ block created by $p$. 
    \item \textbf{Member-Created Feeds}: The feed of $p$ includes the $(\textsc{say},x)$ and  $(\textsc{respond},x,\textit{id})$ blocks created by $p$.
    \item \textbf{Grassroots Dissemination}:
     $p$-blocks sent to friends include the identities  (signed hash pointers) of the tips of $B$, and
     $p$ eventually sends every friend $q$ every $q'$-block $b \in B$ not observed by any $q$-block in $B$, for which $B$ includes a $(\textsc{follow}, q')$ $q$-block. 
\end{enumerate}
   }
    }


 \vspace{0.5em}
\begin{figure}[tp]
\begin{algorithm}[H]
	\caption{\textbf{\TL: Twitter-Like Protocol}\\ Code for agent $p$, including Algorithm \ref{alg:blocklace}}	\label{alg:tl}
	\small
	\begin{algorithmic}[1] \scalefont{0.93}
	\alglinenoPop{counter} 
     \Statex \textbf{Local variables:}
   \State B $\gets \{\}$  \Comment{The local blocklace of agent $p$}  \label{tl:blocklace}

\vspace{0.5em}
	\Upon{decision to utter $x$}   \label{tl:utter}     \Comment{Create a \textsc{follow}, \textsc{say}, or \textsc{respond} block} 
          \State $B\gets B \cup \{\textit{new\_block}(x, \textit{tips}(B))\}$  \label{alg:BDA-receive}
	 \EndUpon

\vspace{0.5em}
	\Upon{change of IP address}   \label{tl:IP-change}     \Comment{Announce new IP address}
          \State $B\gets B \cup \{\textit{new\_block}(\bot, \textit{tips}(B))\}$  \label{alg:BDA-receive}
	 \EndUpon

\vspace{0.5em}

	\Upon{$\textbf{receive } b$}   \label{tl:receive_and_ack}     \Comment{Receive and \textsc{ack} a block} 
            \State $B\gets B \cup \{b\}$  
            \If{$b.\textit{payload}\ne \textsc{ack}$}                   \Comment{No \textsc{ack} of an  \textsc{ack} block} 
            \State \textbf{send} $\textit{new\_block}(\textsc{ack},\textit{ack\_pointers}(b,B))$ to $\textit{ip\_address}(b.\textit{creator},B)$  
            \EndIf
	 \EndUpon

	   	 \vspace{0.5em}	
     \Upon{new block in $B$}    \Comment{Grassroots dissemination/friendship offer}
      \label{tl:grassroots-dissemination}
        \For{ $\forall b\in B~ q\in \Pi :$
        \State $(b.\textit{payload} \ne  \textsc{ack} \wedge \textit{friends}(q, B) \wedge \textit{follows}(q, b.\textit{creator}, B)~ \vee$      \Comment{The friend $q$ follows the creator of $b$}               
        \State $b.\textit{payload}= (\textsc{follow},q))~ \wedge$ \label{tl:frienship-offer} \Comment{Friendship offer to $q$}
        \State $\lnot\textit{aObserves}(q, b, B)$}
	    \State \textbf{send}  $b$ to  $\textit{ip\_address}(q,B)$   \label{alg:send-package}
	    \EndFor
	    \EndUpon

\vspace{0.5em} 

	 \Procedure{\textit{follows}}{$q, q', B$} \label{tl:follows}
            \State \Return $ q= q' \vee \exists b \in B: 
      b.\textit{creator}=q 
     \wedge  b.\textit{payload}=(\textsc{follow},q')$  
        \EndProcedure

\vspace{0.5em} 

	 \Procedure{\textit{friends}}{$q, B$} \label{tl:friends}
            \Return $\textit{follows}(p, q, B) \wedge \textit{follows}(q, p, B)$  
        \EndProcedure

 		\vspace{0.5em}
		\Procedure{$\textit{ack\_pointers}$}{$b,B$}   
        \If{$\textit{friend}(b.\textit{creator}, B)$} \Return $\textit{tips}(B)$  \Comment{Disclose what you know to a friend} \label{tl:disclose}
        \EndIf
        \If{$b.\textit{payload}= (\textsc{follow},p)$} \Return $\{b.\textit{id}\}$  \Comment{Ack an offer} \label{tl:ack-follow}
        \EndIf
        \EndProcedure
 
		\alglinenoPush{counter}
	\end{algorithmic}

\end{algorithm}
\vspace{-2em}
\end{figure}

\mypara{Algorithm \ref{alg:tl} walkthrough} Pseudocode specifying the protocol is presented as Algorithm \ref{alg:tl}.  Here is a walk through the pseudocode for agent $p$:  The agent maintains a local blocklace $B$ that is initially empty (\Cref{tl:blocklace}).
The agent may \textsc{follow} another agent, \textsc{say} something to their personal feed (`tweet'), or \textsc{respond} to something that was said, by adding a new block $b$ with the corresponding payload to the blocklace (\Cref{tl:utter}).
Upon change of an IP address, the agent $p$ issues a new block with the updated IP address (\Cref{tl:IP-change}).
Upon receipt of a new block $b$, the agent $p$ adds $b$ to the blocklace $B$, creates an \textsc{ack}-block, and sends it to the creator of $b$ without storing the \textsc{ack}-block in its blocklace (\Cref{tl:receive_and_ack}).  

Once the agent adds a new block to the blocklace (whether received or created), grassroots dissemination is activated (\Cref{tl:grassroots-dissemination}), upon which $p$ sends to every agent $q$ every block needed by $q$ (e.g., a block by an agent $q'$ followed by $q$), but not known to $q$ according to the $q$-blocks in $p$'s blocklace.  This code also takes care of sending and resending to $q$ a friendship offer, namely a $(\textsc{follow},q)$ block, until acknowledged (\Cref{tl:frienship-offer}).  Acknowledgment of receipt of the offer does not imply acceptance --  to accept, $q$ has to create the corresponding $(\textsc{follow},p)$ block (\Cref{tl:utter}).

The protocol uses procedures for checking that $p$ follows $q$ (\Cref{tl:follows}) and that $p$ and $q$ are friends, namely follow each other (\Cref{tl:friends}).  It also uses a procedure to determine the hash pointers in an \textsc{ack}-block:  Fully-disclose to a friend the current tips of the blocklace (\Cref{tl:disclose}) or just acknowledge receipt of a friendship offer (\Cref{tl:ack-follow}).  This completes the walkthrough of Algorithm \ref{alg:tl}.

Next we consider the security and spam-resistance of the \TL protocol.

\mypara{Safety} The safety requirement of GSN is that any utterance of a member can be attributed to that member, and in the correct context in which it was uttered.  Any \TL \textsc{say} and \textsc{respond} block $b$ is signed by its creator and  includes hash pointers to the blocks that are causally-precedent to $b$.   Thus, such an utterance  can be attributed to its author and  be forwarded with correct attribution and context, and if forwarded multiple times would  carry their entire provenance.

\mypara{Liveness}
The \TL liveness condition is that a follower may receive any utterance of an author it follows if the follower is connected in the social graph to the author via a path of correct members all of which follow the author.  A similar liveness theorem was proven in reference~\cite{shapiro2023grassroots}.  The intuition is that any utterance in a block $b$ of the follower will eventually traverse each edge $p\leftrightarrow q$.  Assuming $p$ already knows $b$, then since $q$ also follows the author of $b$ and $p$ and $q$ are friends then $p$ will eventually know that $q$ follows the author of $b$. And if $q$ does not know $b$ then $p$ will eventually know that $q$ does not know $b$, and hence $p$ will eventually send $b$ to $q$, which will eventually received $b$.  And so forth for each edge in the path.

\mypara{Privacy} \TL is a public network so no need for encryption and any member may follow the  feed of any other member.

\mypara{Spam, Bots and Deep Fake}  Each utterance is signed by its author and every forwarded block has provenance, so that anything without acceptable origin or provenance can be safety ignored.

\section{\WL: A Grassroots WhatsApp-Like Protocol}\label{section:WL}

Here we present a WhatsApp-like protocol termed \WL, within the Grassroots Social Networking architecture.  The protocol employs blocks with the following payload keywords:
\begin{enumerate}
    \item \textsc{group}, to create a new group/hyperedge in the social graph,
      \item \textsc{invite}, to invite a member to join a group,\footnote{The handling of \textsc{remove} and \textsc{leave} is deferred; it is not conceptually difficult but will clutter the exposition.}
     \item \textsc{accept}, to accept an invitation to join a group,
    \item \textsc{say}, to post to a group,  
    \item \textsc{respond}, to respond to a post or a response, and
    \item \textsc{ack}, to acknowledge receipt of a block and disclose the group blocks one knows
\end{enumerate}  

In  \WL, an agent $p$ founds a new group, with itself as the sole member,  by creating a \emph{group genesis} initial block of the form $(\textit{id},\textit{address},(\textsc{group},\textit{name}),\emptyset)$, where \textit{name} is an arbitrary string (e.g., it could encode the group's name, description, and photo) that has not been previously used by $p$ to create a group. Recall that in this case \textit{id} is $\textit{hash}((\textit{address},(\textsc{group},\textit{name}),\emptyset))$ signed by $p$.

A correct \WL block observes exactly one group genesis block and as such pertains to that group, and a correct agent maintains a closed blocklace.  Therefore, the blocklace $B$ of a correct agent $p$ participating in \WL can be partitioned into disjoint closed subsets, with partition $B_\textit{id}$ including the blocks that observe the group genesis block \textit{id}.  Intuitively, the partitions of the blocklace of $p$ correspond to the groups $p$ is a member of, with each partition including the blocks known to $p$ that pertain to that group.  

The \WL protocol realizes the principles and components of Grassroots Social Networking thus:\\
\newline
\fcolorbox{gray!100}{gray!5}{%
    \parbox{0.96\textwidth}{%
\textbf{\WL Realization of \GSN}\\ by each agent $p$ with blocklace $B$ partitioned into groups, with $B_{\textit{id}}$ being the \textit{id}  partition of $B$
\begin{enumerate}[leftmargin=*]
    \item \textbf{Friendship}: $p$ joins a group \textit{id} created by $q$ by accepting an invitation from $q$.  $q$ invites $p$ to \textit{id} by creating the block $(\textit{id}',\textit{address},(\textsc{invite},q),\{\textit{id}\})$ and $p$ accepts the invitation by creating the block $(\textit{id}'',\textit{address}',\textsc{accept},\{\textit{id}'\})$.
 
    \item \textbf{Decentralized Social Graph}: A new hyperedge labeled \textit{id} with $p$ as its founder and sole vertex is added to the social graph by $p$ creating a group genesis block $(\textit{id},\textit{address},(\textsc{group},\textit{name}),\emptyset)$.  A vertex $q$ is added to a hyperedge labeled \textit{id} with founder $p$ by $q$ accepting an invitation from $p$ to join the group  \textit{id}.
 
    \item \textbf{Member-Created Feeds}: Each group \textit{id} has an associated feed, with group members as authors and followers and with content being member-created $(\textsc{say},x)$ and  $(\textsc{respond},x,\textit{id})$ blocks that observe \textit{id}.
   
    \item \textbf{Grassroots Dissemination}:
     $p$-blocks sent to group members include the identities  (signed hash pointers) of the tips of the group's partition in $B$.
     For every group of which $p, q, q'$ are members, $p$ eventually sends $q$ every $q'$-block not observed by any $q$-block in $B$.
\end{enumerate}
   }
    }
    \vspace{0.5em}

\begin{figure}[tp]
\begin{algorithm}[H]
	\caption{\textbf{\WL: WhatsApp-Like Protocol}\\ Code for agent $p$, including Algorithm \ref{alg:blocklace}}	\label{alg:wl}
	\small
	\begin{algorithmic}[1] \scalefont{0.93}
	\alglinenoPop{counter} 
		
    \Statex \textbf{Local variables:}
    \State B $\gets \{\}$  \Comment{The local blocklace of agent $p$}  \label{wl:blocklace}

\vspace{0.5em}
	\Upon{decision to create a new group named \textit{name}}   \label{wl:create}
        \State $b \gets \textit{new\_block}((\textsc{group},\textit{name}),\emptyset)$ 
	 \EndUpon

\vspace{0.5em}
	\Upon{decision to invite $q$ to group \textit{id}}   \label{wl:invite}   
        \State $b \gets \textit{new\_block}((\textsc{invite},q),\{\textit{id}\})$
	 \EndUpon

\vspace{0.5em}
	\Upon{decision to accept invitation to join group \textit{id}}   \label{wl:accept}   
        \If {$\exists b= (\textit{id}',(\textsc{invite},p),\{\textit{id}\}) \in B  \wedge
                \textit{id}.\textit{creator} =  \textit{id}'.\textit{creator}$}
         $b \gets \textit{new\_block}(\textsc{accept},\{\textit{id}'\})$
        \EndIf
	 \EndUpon

\vspace{0.5em}
	\Upon{decision to say $x$ to the group of block \textit{id}}   \label{wl:say}  
        \State $b \gets \textit{new\_block}((\textsc{say},x),\textit{tips}(\textit{group}(B,\textit{id})))$
	 \EndUpon

\vspace{0.5em}
	\Upon{decision to respond $x$ to block \textit{id}}   \label{wl:respond}   
        \State $b \gets \textit{new\_block}((\textsc{respond},x,\textit{id}),\textit{tips}(\textit{group}(B,\textit{id})))$
	 \EndUpon

\vspace{0.5em}
	\Upon{change of IP address}   \label{wl:IP-change}     \Comment{Announce new IP address}
          \For{$\forall \textit{id}: (\textit{id},(\textsc{group},\textit{name}),\emptyset) \in B$} 
          \State $B\gets B \cup \{\textit{new\_block}(\bot, \textit{tips}(\textit{group}(B,\textit{id})))\}$ 
          \EndFor \label{alg:BDA-receive}
	 \EndUpon

\vspace{0.5em}

	\Upon{$\textbf{receive } b$}   \label{wl:receive_and_ack}     \Comment{Receive and \textsc{ack} a block} 
            \State $B\gets B \cup \{b\}$  
            \If{$b.\textit{payload}\ne \textsc{ack}$}                   \Comment{No \textsc{ack} of an  \textsc{ack} block} 
            \State \textbf{send} $\textit{new\_block}(\textsc{ack},\textit{ack\_pointers}(b,B))$ to $\textit{ip\_address}(b.\textit{creator},B)$  
            \EndIf
	 \EndUpon

	   	 \vspace{0.5em}	
     \Upon{new block $b'\in B$}    \Comment{Grassroots dissemination/invitation}
      \label{wl:grassroots-dissemination}
        \For{ $\forall b\in B~ q\in \Pi :$ 
        \State $(b.\textit{payload} \ne  \textsc{ack}\wedge \textit{member}(q, b'.\textit{id})~ \vee$  
        \State $b.\textit{payload}= (\textsc{invite},q))~ \wedge$ \label{wl:send-invite} \Comment{Invitation}
        \State $\lnot\textit{aObserves}(q, b, B)$}
	    \State \textbf{send}  $b$ to  $\textit{ip\_address}(q,B)$   \label{alg:send-package}
	    \EndFor
	    \EndUpon

\vspace{0.5em} 

	 \Procedure{\textit{member}}{$q, B$} \label{wl:member}
            \Statex \Comment{$q$ is a member of a group if it is its creator or if it was invited by its creator $q'$ and accepted the invitation}
            \State \Return $ \exists b, b', b'' \in B: $ 
           
            \State $b = (\textit{id},(\textsc{group},\textit{name}),\emptyset) \wedge  \textit{id}.\textit{creator}=q' \wedge$
            \State $(q =q' \vee b' = (\textit{id}',(\textsc{invite},q),\textit{id}) \wedge \textit{id}'.\textit{creator} = q' \wedge$
            \State   $b''=(\textit{id}'', \textsc{accept},\{\textit{id'}\}) \wedge \textit{id}''.\textit{creator} = q)$  
        \EndProcedure

\vspace{0.5em}

 		\vspace{0.5em}
		\Procedure{$\textit{ack\_pointers}$}{$b,B$}   
        \If{$\textit{member}(p, b.\textit{id})$} \Return $\textit{tips}(\textit{group}(B,b.\textit{id}))$  \Comment{Disclose to a member} \label{wl:disclose}
        \EndIf
        \If{$b.\textit{payload}= (\textsc{invite},p)$} \Return $\{b.\textit{id}\}$  \Comment{Ack an invite} \label{wl:ack-invite}
        \EndIf
        \EndProcedure

 		\vspace{0.5em}
		\Procedure{$\textit{group}$}{$B,\textit{id}$}  
        \State \textit{Let} $b=(\textit{id},x,H) \in B, b' = (\textit{id}',(\textsc{group},\textit{name}),\emptyset)$ s.t. $\textit{observes}(b,b')$
        \State \Return $\{b''\in B :  \textit{observes}(b'',b',B) \}$
        \EndProcedure

		\alglinenoPush{counter}
	\end{algorithmic}

\end{algorithm}
\vspace{-2em}
\end{figure}

\mypara{Algorithm \ref{alg:wl} walkthrough} Pseudocode specifying the protocol is presented as Algorithm \ref{alg:wl}.  Here is a walk through the pseudocode for agent $p$:  The agent maintains a local blocklace $B$ that is initially empty (\Cref{wl:blocklace}).
The agent $p$ may create a new \textsc{group} (\Cref{wl:create}); \textsc{invite} another agent to a group $p$ created (\Cref{wl:invite}); \textsc{accept} in invitation to group and thus join it (\Cref{wl:accept});  \textsc{say} something to a group one is a member of (\Cref{wl:say}); or \textsc{respond} to something that was said (\Cref{wl:respond}).
Upon change of an IP address, the agent $p$ issues a new block with the updated IP address (\Cref{wl:IP-change}).
Upon receipt of a new block, the agent $p$ adds it to the blocklace $B$, creates an \textsc{ack}-block, and sends it the creator of the received block, without storing the \textsc{ack}-block in its blocklace (\Cref{wl:receive_and_ack}).  

Once the agent adds a new block to the blocklace (whether received or created by $p$), grassroots dissemination is activated (\Cref{wl:grassroots-dissemination}), upon which $p$ sends every block $b$ of the group to every member $q$ of that group that does not know $b$ according to the $q$-blocks in $p$'s blocklace.  This code also takes care of sending and resenting to $q$ an \textsc{invite} (\Cref{wl:send-invite}).  Acknowledgment of receipt of the invite does not imply acceptance --  to accept, $q$ has to create the corresponding \textsc{accept} block (\Cref{wl:accept}).

The protocol uses procedures for checking that $q$ is a member of a group (\Cref{wl:member}).  It also uses a procedure to determine the hash pointers an \textsc{ack}-block should carry:  Fully-disclose to a member of the same group (\Cref{wl:disclose}) or just acknowledge receipt of an invitation (\Cref{wl:ack-invite}).  This completes the walkthrough of Algorithm \ref{alg:wl}.

Next we consider the security and spam-resistance of the \WL protocol.

\mypara{Safety} Similarly to \TL,   any \WL \textsc{say} and \textsc{respond} block $b$ is signed by its creator and  includes hash pointers to the blocks causally-precedent to $b$.   Thus, such an utterance  can be attributed to its author and be forwarded with correct attribution and context, and if forwarded multiple times among different groups it would  carry its entire provenance.

\mypara{Liveness}
The \WL liveness condition is that an utterance to a group known to a correct group member will eventually reach every correct group member.   The reason is that if a block $b$ of a group of which both $p$ and $q$ are correct members is known to $p$ and not to $q$, then $p$ will eventually know that $q$ does not know $b$, and hence $p$ will eventually send $b$ to $q$, which will eventually receive $b$. 

\mypara{Privacy} In \WL each group is private.  Privacy is realized by the group founder creating a special keypair for the group and secretly sharing the private group key with every new group member.  When a member is removed from a  group or leaves it, the founder has to renew the group's keypair.  Note that, as always, nothing prevents a group member from breaching privacy by sharing the group's key and/or group messages with others.

Another privacy issue is related to messages forwarded across groups.  The forwarder may decipher the message before forwarding, and send it over with the original message and the group's public key so that message authenticity can be ascertained. However, by doing so the sender may reveal additional information through the block's hash pointers, e.g. the identities of other group members.  As authenticated forwarding may be desirable, it can be achieved without unnecessarily compromising privacy by requiring the payload itself of each block to be signed.

\mypara{Spam and Deep Fake}  As in \TL, each utterance is signed by its author and every forwarded block has provenance.

\section{\GSN and Secure Scuttlebutt}\label{section:scuttlebutt}

The protocol closest to \GSN, in spirit and goal, is 
Secure Scuttlebutt (SSB)~\cite{tarr2019secure,kermarrec2020gossiping,tschudin2019broadcast}, a peer-to-peer protocol, mesh network, and self-hosted social media platform. Unlike \GSN, SSB has been implemented and is being used. 

SSB offers single-authored, authenticated, append-only logs, aka personal blockchains, and a dissemination protocol in which members that follow the same logs update each other when they connect. 
We compare it to the \GSN Twitter-Like protocol.  In terms of data structures, SSB has no social graph along the edges of which communication occurs; thus communication is in principle all-to-all. As a result, agents have to actively block other agents they do not wish to communicate with.
The blocklace created by the Twitter-Like protocol has a `personal blockchain' structure similar to SSB logs, but in addition has pointers to other agents' `personal blockchains' the agent follows, indicating the most recently-known blocks of the followed agents.  Agents communicate only with their friends, which in the Twitter-Like protocol are agents that voluntarily follow each other, so no need for blocking.

In terms of dissemination, the SSB protocol is connection-based:  When two peers connect they exchange information to synchronize their shared logs. It is geared for peers that know each other's IP address and use the information to establish reliable communication links (TCP), namely networked computers.  Grassroots Dissemination is connectionless:  Every so often, an agent $p$ sends to every friend $q$ every block $p$ knows and believes that $q$ needs, based on the last block received from $q$.  Furthermore, each block sent includes the most recent IP address of the sender, allowing agents to keep track of their friend's changing IP addresses. Thus it is geared for peers with dynamically-changing IP addresses that communicate via unreliable messages (UDP), namely smartphones.

More generally, the notions of \GSN---the blocklace, the social graph, and grassroots dissemination--- 
are more abstract and may be more broadly applicable than SSB, with the Twitter-like protocol,  the WhatsApp-like protocol and the implementation of of grassroots cryptocurrencies~\cite{lewis2023grassroots}, being but examples.

\section{Conclusions}\label{section:conclusions}
We have presented the architecture of Grassroots Social Networking with two protocols -- Twitter-Like and WhatsApp-like.   Subsequent work would include latency-optimizing variants and extensions of these protocols, develop a prototype implementation of the protocols presented here and the protocols for grassroots currencies~\cite{shapiro2022gc,lewis2023grassroots}, and additional applications of the architecture presented in~\cite{shapiro2024grassroots}.

\begin{acks}
I thank Andrew Lewis-Pye, Oded Naor, Suhail Manzoor, Jacob Bassiri, and Erez Kedem for discussions and feedback.
Ehud Shapiro is the Incumbent of The Harry Weinrebe Professorial Chair of Computer Science and Biology at the Weizmann Institute and a Visiting Professor at the London School of Economics.
\end{acks}
\newpage

\bibliographystyle{ACM-Reference-Format}
\bibliography{bib}

\end{document}